# A SIMULATION MODEL FOR THE LIFE-TIME OF WIRELESS SENSOR NETWORKS

Abdelrahman Elleithy [1] and Gonhsin Liu [2]

[1] Department of Computer Science and Engineering, University of Bridgeport, Bridgeport, CT, USA
aelleith@bridgeport.edu
[2] Department of Computer Science and Engineering, University of Bridgeport, Bridgeport, CT, USA
gonhsin@bridgeport.edu

## ABSTRACT

*In this paper we present a model for the lifetime of wireless sensor networks. The model takes into consideration several parameters such as the total number of sensors, network size, percentage of sink nodes, location of sensors, the mobility of sensors, and power consumption. A definition of the life time of the network based on three different criteria is introduced; percentage of available power to total power, percentage of alive sensors to total sensors, and percentage of alive sink sensors to total sink sensors. A Matlab based simulator is developed for the introduced model. A number of wireless sensor networks scenarios are presented and discussed.*

## 1. A MODEL FOR LIFETIME OF WIRELESS SENSORS NETWORK

Wireless sensors have received increased attention in the past years due to their popularity and cost effectiveness when they are used in harsh environments. They have been used in many applications including military applications, environmental applications, health applications, and home applications. Although they are very cost effective and easily deployed in harsh environments, they are limited by the power available through their life cycle. Sensors are usually deployed with limited power which is depleted over their lifecycle. Once their power is depleted, the sensors become dead and they are no more useful. An evaluation of the life cycle of a wireless sensor network is very essential to estimate how long a network can live and when the network and its sensors might be replaced or recharged if possible.

In this section we present a model for the lifetime of Wireless sensor networks based on a paper by [1]. The model takes different parameters that used in literature. The following parameters are considered:

1. The time until the first sensor is drained of its energy [2];

2. The time until the first cluster head is drained of its energy [3] ;

3. The time there is at least a certain fraction β of surviving nodes in the network [4];

4. The time until all nodes have been drained of their energy [5];

5. K-coverage: the time the area of interest is covered by at least k nodes [6] ;

6. 100% coverage

   a. The time each target is covered by at least one node [7] ;

   b. The time the whole area is covered by at least one node [8] ;

7. α-coverage

   a. The accumulated time during which at least α portion of the region is covered by at least one node [9];

   b. The time until the coverage drops below a predefined threshold α (until last drop below threshold) [10] ;

   c. The continuous operational time of the system before either the coverage or delivery ratio first drops below a predefined threshold [11];

8. The number of successful data-gathering trips [12] ;

9. The number of total transmitted messages [13];

10. The percentage of nodes that have a path to the base station [11];

11. Expectation of the entire interval during which the probability of guaranteeing connectivity and k-coverage simultaneously is at least α [6];

12. The time until connectivity or coverage are lost [14];

13. The time until the network no longer provides an acceptable event detection ratio [5];

14. The time period during which the network continuously satisfies the application requirement [15];

15. min(t1, t2, t3) with t1: time for cardinality of largest connected component of communication graph to drop below c1 × n(t), t2: time for n(t) to drop below c2 ×n, t3: time for the covered volume to drop below c3 ×l d [16].

## 2. PARAMETERS USED IN THE MODEL

In this section we address parameters that were introduced in literature that can be used in a complete model for a wireless sensors networks life time. The following parameters are introduced:

1. The total number of available sensors

2. The set of all nodes those that are alive at a certain time t

3. The set of nodes those that are active at a time t

4. The set of nodes those that are active at any time in the time interval [t −_t, t]

5. The set of sink nodes or base stations B(t) is defined to be a subset of the existing nodes SY

6. The ability of nodes m1 and mn to communicate at a time t

7. The ability of two nodes to communicate in the time interval [t −_t, t] such that the links between consecutive hops become available successively within the time interval (support for delay tolerant networking)

8. The set of target points to be sensed by the network

9. The area that is covered by all sensors of a certain type y, at a time t.

## 3. DEFINITION OF NETWORK LIFETIME

There are two network lifetime metrics that introduced in literature based on definitions on the previous sections. Both metrics depict the network lifetime in seconds. The metrics probably become most expressive when used together.

(1) The first metric gives the accumulated network lifetime Za as the sum of all times that $\zeta$ (t) is fulfilled, stopping only when the criterion is not fulfilled for longer than $\Delta$ tsd seconds.

(2) The second metric, the total network lifetime Zt , gives the first point in time when the liveliness criterion is lost for a longer period than the service disruption tolerance $\Delta$ tsd.

## 4. SIMULATION OF THE LIFETIME OF WSN

In this section we present the following model for simulating the life time in a WSN. Although the Matlab code is developed with the following default parameters, it can be modified for other values.

Assumptions:

1. Total number of sensors:          Input parameter

2. Network size:              Input parameter in meters

3. Percentage of sink sensors:        Input parameter (between 0 and 1)

4. Location of sensors:              Randomly generated over the network

5. Initial power per sensor:  Random between 0 and 100 units

6. Movement of sensors:

    a)    sensors can move in the x direction for a random value between -5 and +5

    b)    sensors can move in the y direction for a random value between -5 and +5

    c)    The total value a senor moves is :

7. Power Consumption:

    a.  Communications:     1 unit for regular sensor and 2 units for sink sensor

    b.  Movement:         d unit for the moving sensor as calculated in 6

8. Stopping criteria (we consider the network dead if one of the following conditions satisfied):

    a. Percentage of available power to total power:  less than 25 %

    b. Percentage of alive sensors to total sensors:    less than 25 %

    c. Percentage of alive sink sensors to total sink sensors: less than 5 %

## 4.1. Scenario Number 1

In this scenario the following parameters are considered:

1. Total number of sensors:      150

2. Network size:             3000 m X 3000 m

3. Percentage of sink sensors:   15.6 %

Figure 1 shows the status of Initial and final network.

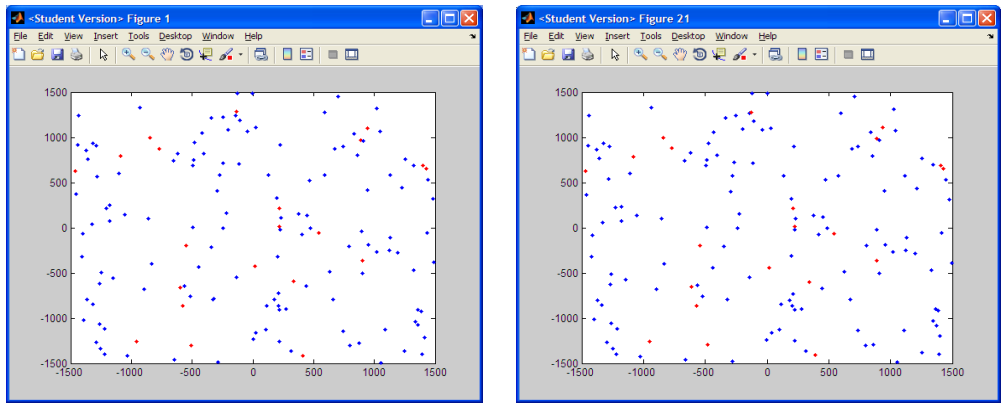

Note: sink nodes are red

Figure 1: Status of Initial and final network.

Figures 2 to 5 show different network performance parameters over the life cycle of the network. In this scenario the network was considered dead because after 21 cycles the following status is reached:

1. Percentage of available power to total power:     1810 / 7751 =   23 %

2. Percentage of alive sensors to total sensors:     78 / 150    =   52 %

3. Percentage of alive sink sensors to total sink sensors:  12 / 21     =   57 %

The network is considered dead because that condition number (1) is satisfied where the percentage of available power to total power is 23 % which is less than the 25 % criteria.

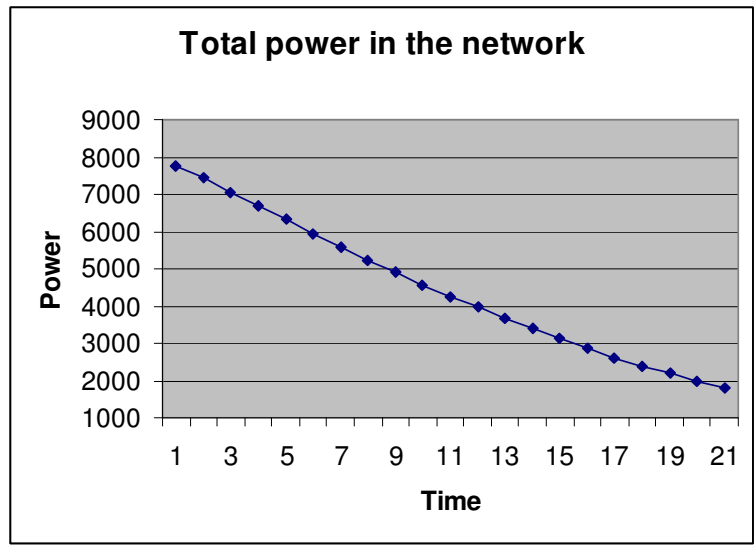

Figure 2: Total Power in the network over the life cycle

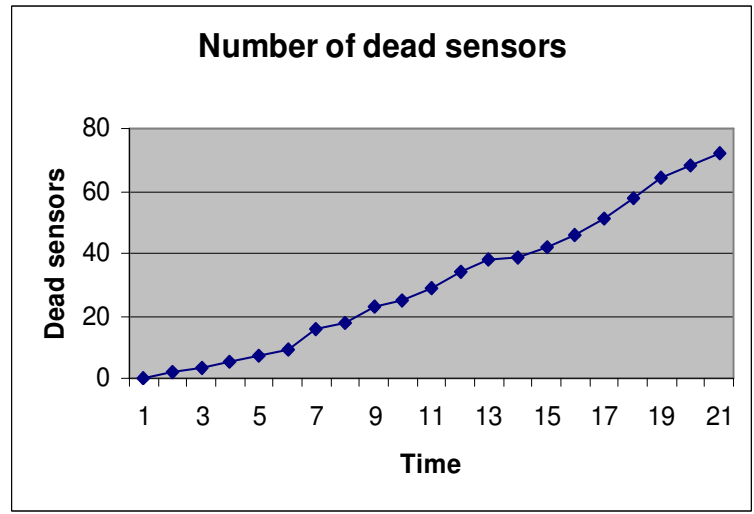

Figure 3: Numbers of dead sensors over the life cycle

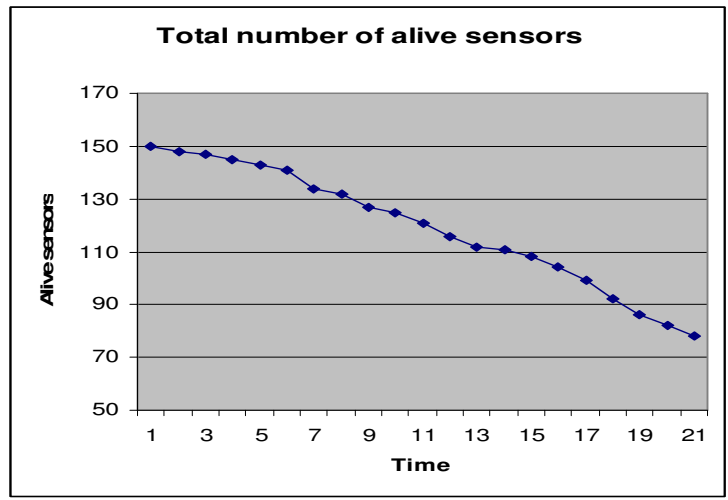

Figure 4: Total number of alive sensors over the life cycle

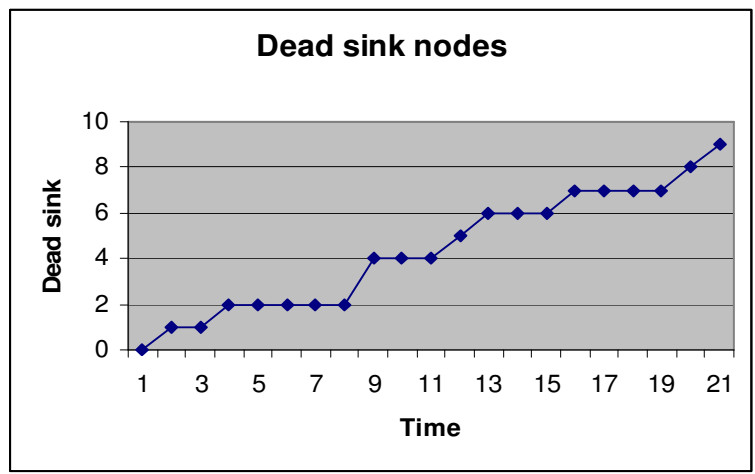

Figure 5: Total number of dead sinks nodes over the life cycle

## 4.2. Scenario Number 2

In this scenario the following parameters are considered:

1. Total number of sensors:      100

2. Network size:                 2000 m X 2000 m

3. Percentage of sink sensors:   10.0 %

Figure 6 shows the status of Initial and final network.

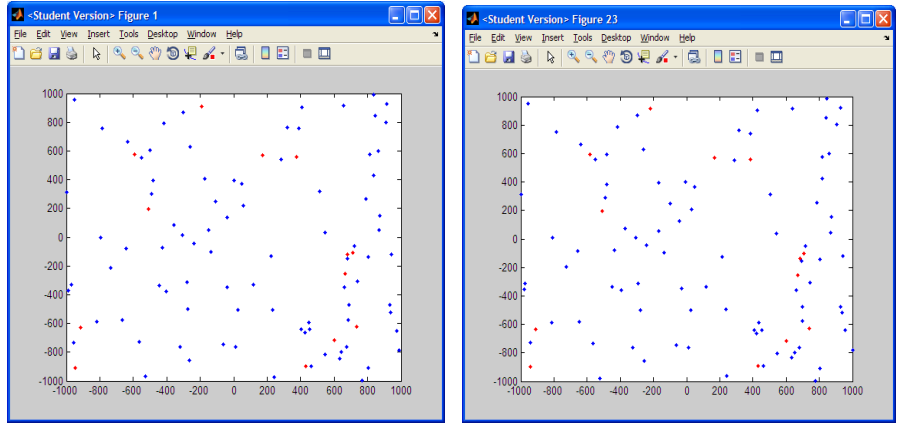

Note: sink nodes are red.

Figure 6:  shows the status of Initial and final network.

Figures 7 to 10 show different network performance parameters over the life cycle of the network. In this scenario the network was considered dead because after 23 cycles the following status is reached:

1.    Percentage of available power to total power:    1248. / 5352   =        23 %

2.    Percentage of alive sensors to total sensors:    50 / 100        =        50 %

3.    Percentage of alive sink sensors to total sink sensors:  5 / 13    =        38 %

The network is considered dead because that condition number (1) is satisfied where the percentage of available power to total power is 23 % which is less than the 25 % criteria.

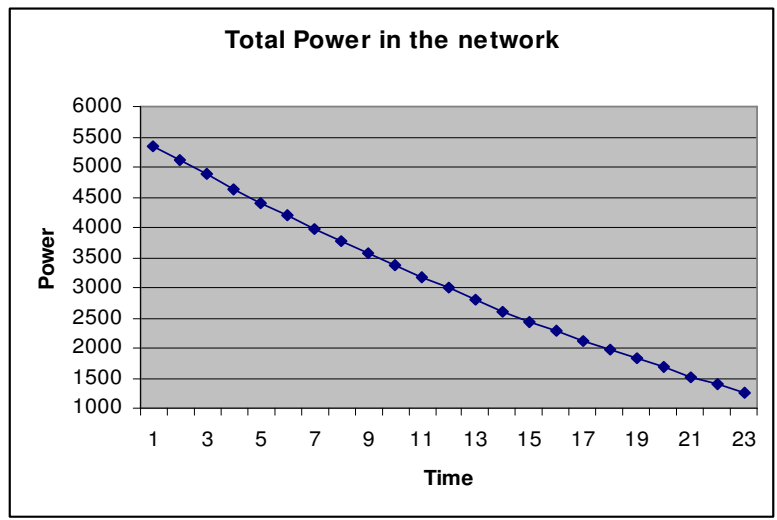

Figure 7: Total Power in the network over the life cycle

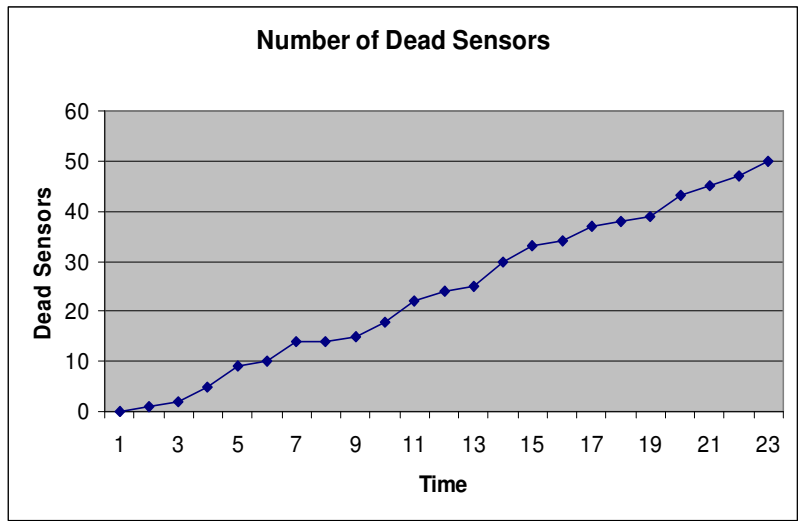

Figure 8: Numbers of dead sensors over the life cycle

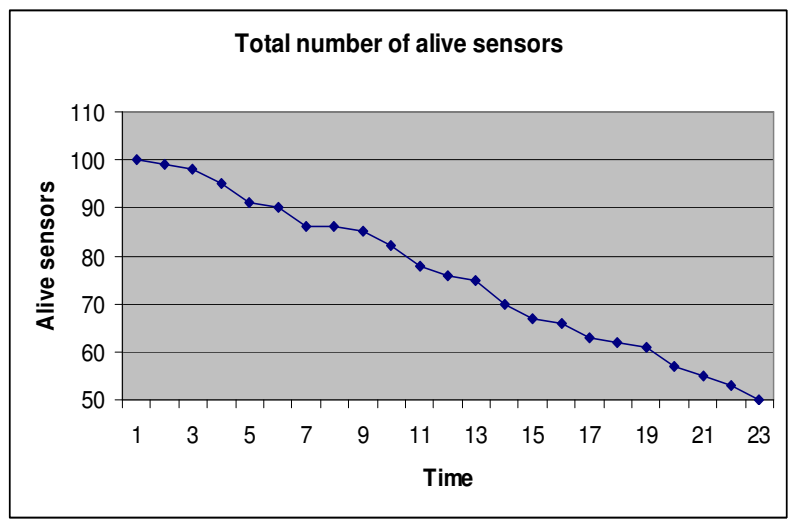

Figure 9: Total number of alive sensors over the life cycle

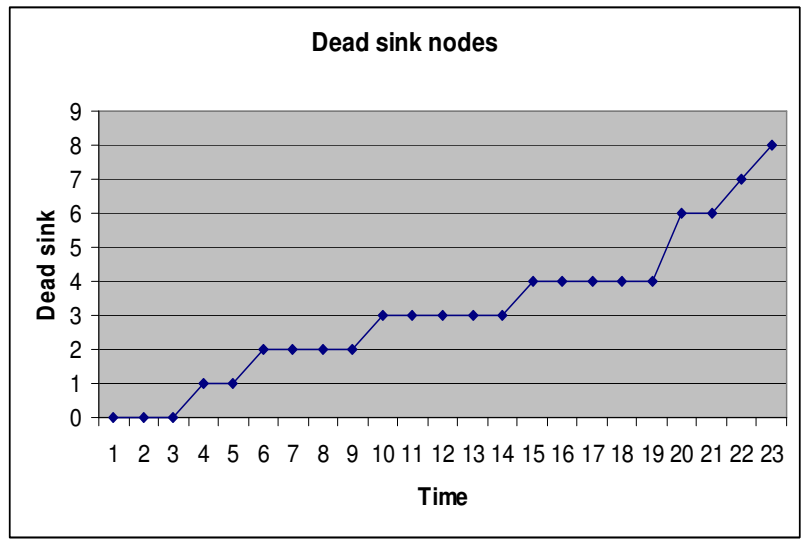

Figure 10: Total number of dead sinks nodes over the life cycle

### 4.3 Scenario Number 3

In this scenario the following parameters are considered:

1.  Total number of sensors:        50

2.  Network size:                    1000 m X 1000 m

3.  Percentage of sink sensors:     20.0 %

Figure 11 shows the status of Initial and final network.

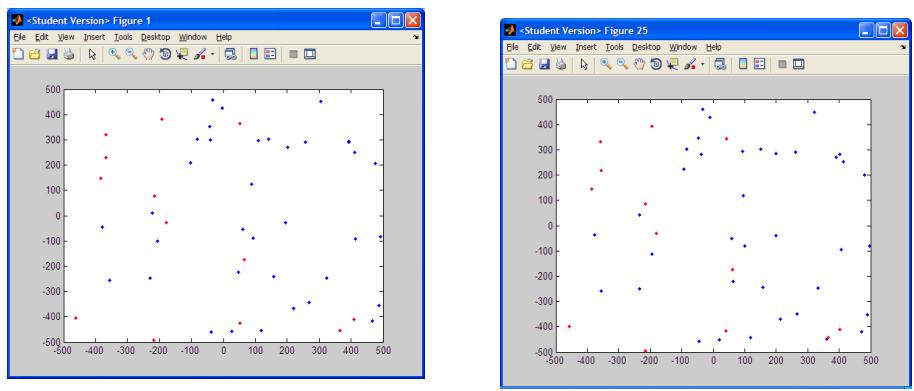

Note: sink nodes are red.

Figure 11: Status of Initial and final network.

Figures 12 to 15 show different network performance parameters over the life cycle of the network. In this scenario the network was considered dead because after 25 cycles the following status is reached:

1. Percentage of available power to total power:   606. / 2542  =  23 %

2. Percentage of alive sensors to total sensors:   23 / 50   = 46 %

3. Percentage of alive sink sensors to total sink sensors:  4 / 13   = 30 %

The network is considered dead because that condition number (1) is satisfied where the percentage of available power to total power is 23 % which is less than the 25 % criteria.

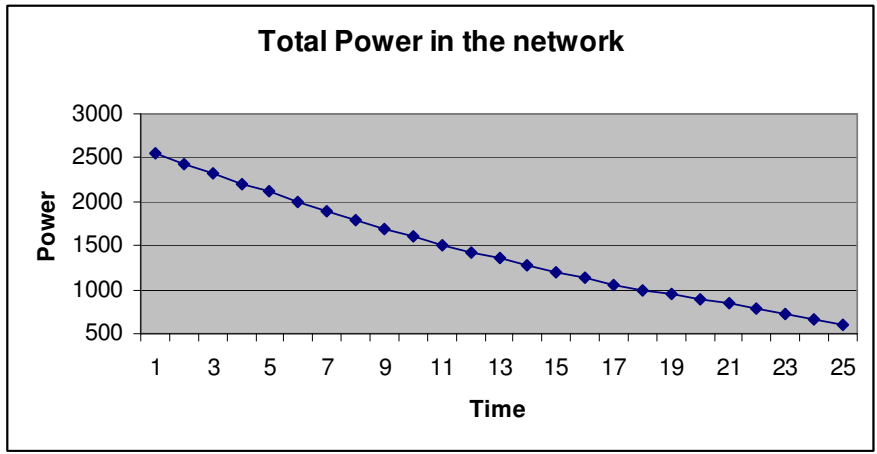

Figure 12: Total Power in the network over the life cycle

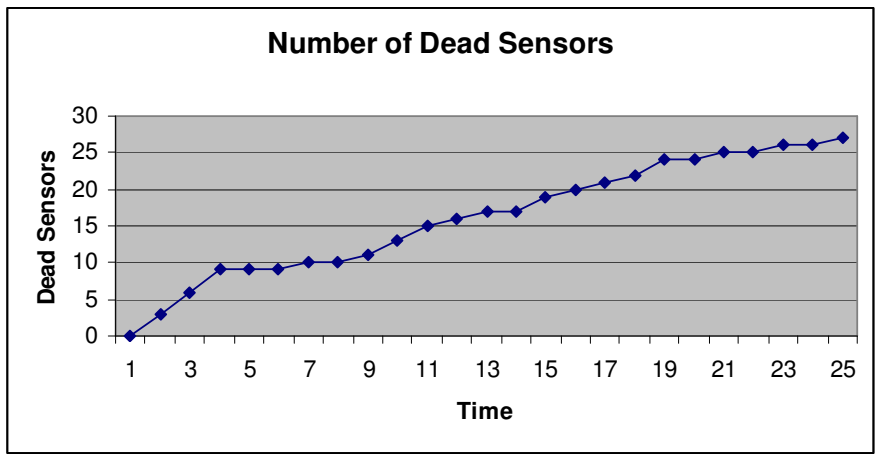

Figure 13: Numbers of dead sensors over the life cycle

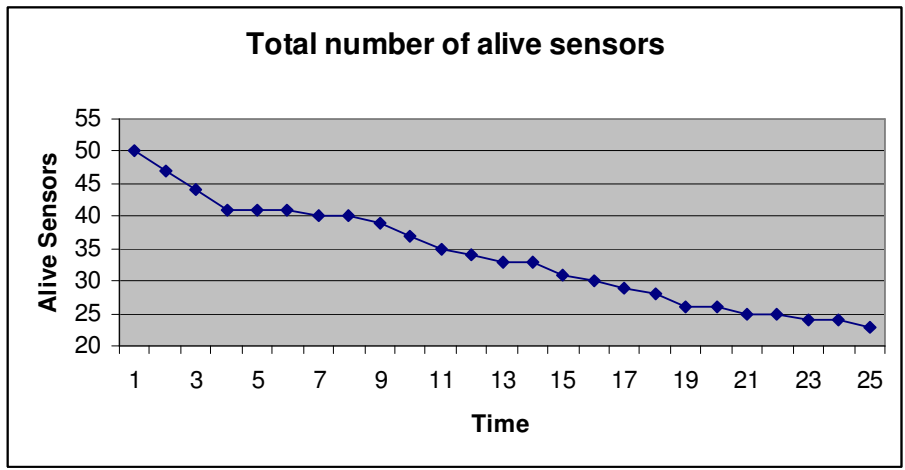

Figure 14: Total number of alive sensors over the life cycle

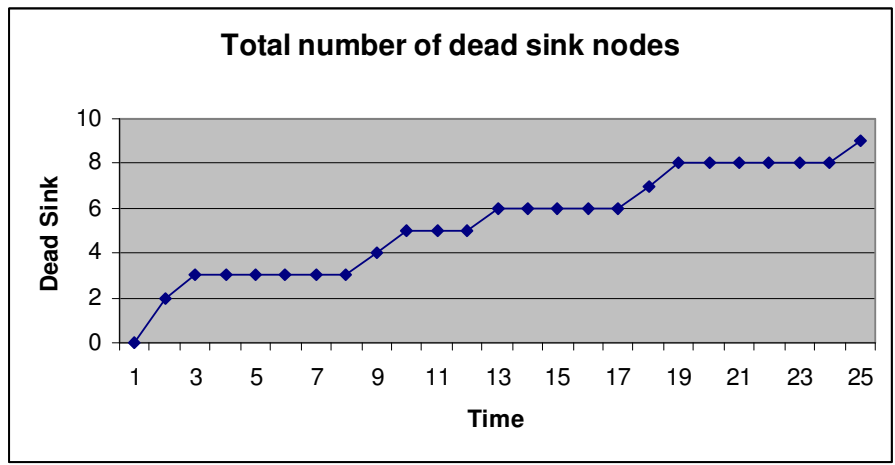

Figure 15: Total number of dead sinks nodes over the life cycle

## 4.4 Scenario Number 4

In this scenario the following parameters are considered:

1. Total number of sensors:               50

2. Network size:               1500 m X 1500 m

3. Percentage of sink sensors:               20.0 %

Figure 16 shows the status of Initial and final network.

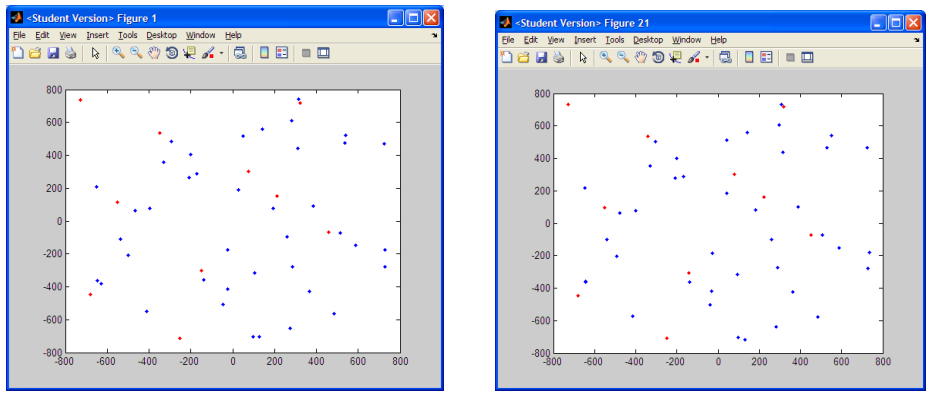

Note: sink nodes are red.

Figure 16: Status of Initial and final network.

Figures 17 to 20 show different network performance parameters over the life cycle of the network. In this scenario the network was considered dead because after 21 cycles the following status is reached:

1. Percentage of available power to total power:       506 / 2069  =  24 %

2. Percentage of alive sensors to total sensors:       20 / 50    = 40 %

3. Percentage of alive sink sensors to total sink sensors:  2 / 10     = 20 %

The network is considered dead because that condition number (1) is satisfied where the percentage of available power to total power is 23 % which is less than the 25 % criteria.

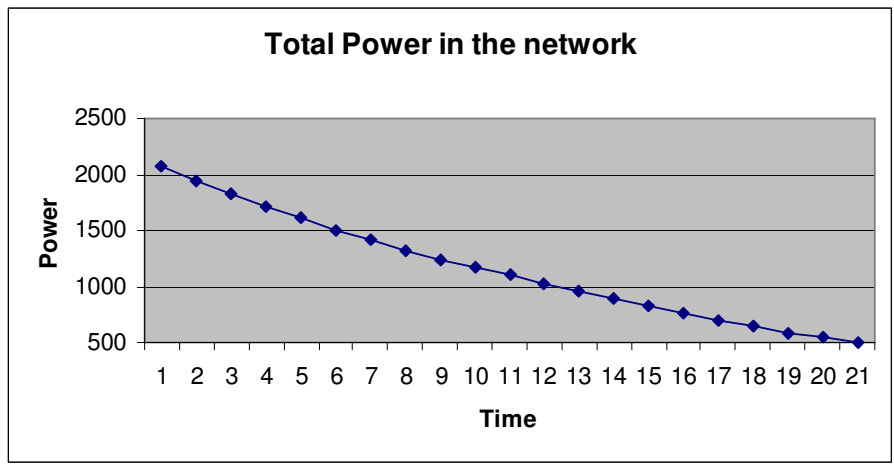

Figure 17: Total Power in the network over the life cycle

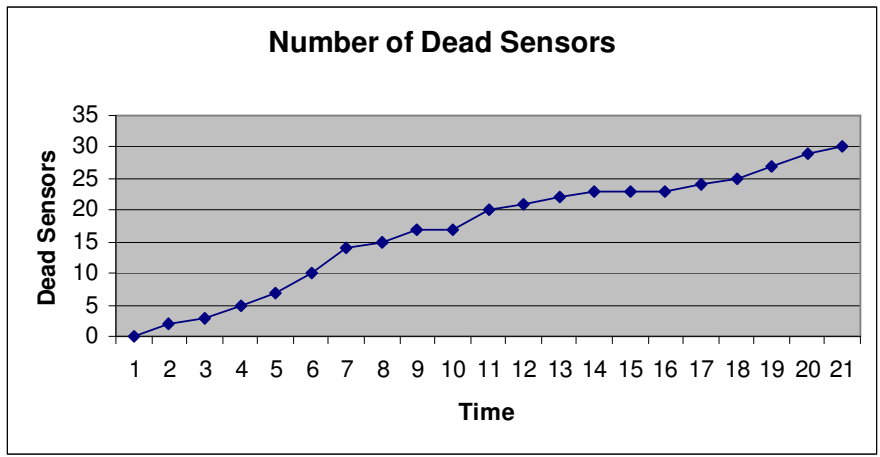

Figure 18: Numbers of dead sensors over the life cycle

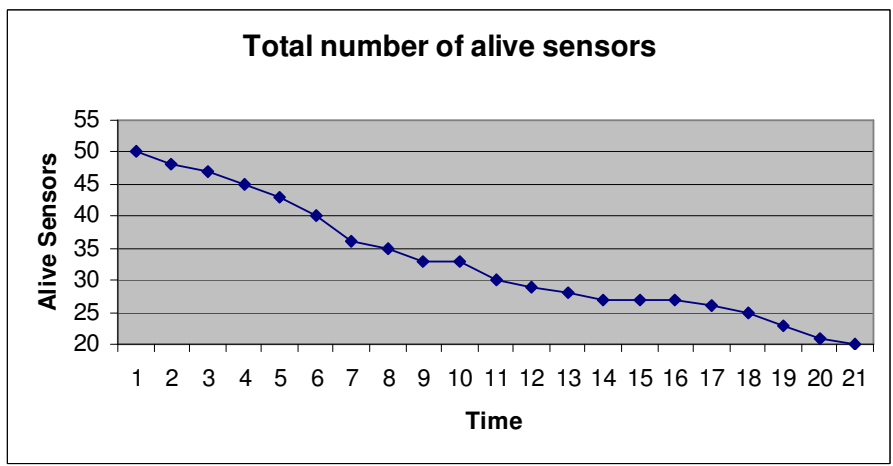

Figure 19: Total number of alive sensors over the life cycle

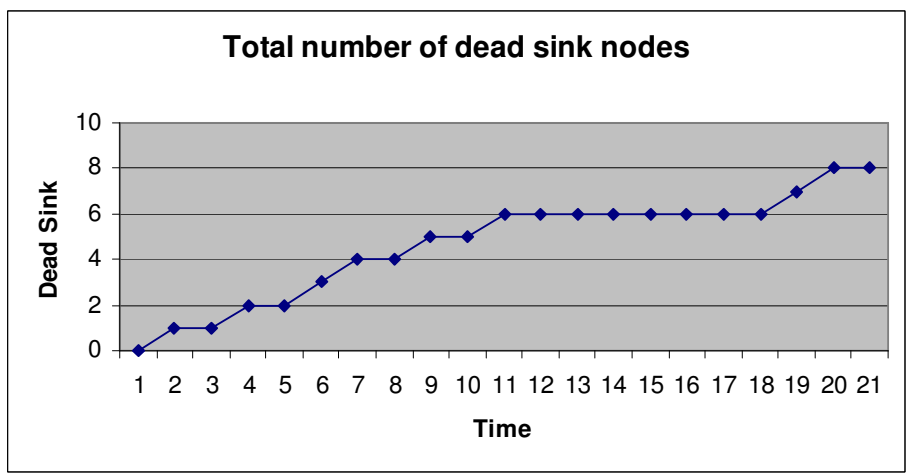

Figure 20: Total number of dead sinks nodes over the life cycle

# 5. CONCLUSIONS

Although wireless sensors networks are very popular, they are limited by the power available through their life cycle. Once their power is depleted, sensors might be replaced or recharged if possible. A model to estimate the life time of wireless sensors networks are an important tool that can help the designers of the network to design their network by adjusting important parameters such as initial power, number of sensors, number of sink sensors, etc.

In this paper we present a model for the lifetime of wireless sensor networks. The model takes into consideration several parameters such as the total number of sensors, network size, percentage of sink nodes, location of sensors, the mobility of sensors, and power consumption. A number of wireless sensor networks scenarios are presented and discussed.

In order to examine the validity of our simulator, we have tested it for many scenarios. In this paper we are presenting four of these scenarios. The following parameters are used: total number of sensors, network size as defined by its width and length, and the percentage of sink sensors. In each scenario, we have evaluated both the total power in the network over the life cycle, number of dead sensors over the life cycle, total number of alive sensors over the life cycle, and number of dead sinks nodes over the life cycle.

The results presented in this paper show how important such a simulator from the designer perspective. The model can be used as a design tool as well as a research tool to evaluate the network performance. We would expect to extend in the future the work presented in this model to include other parameters and other models for defining the life cycle of the wireless sensor networks.

## Authors

**Abdelrahman Elleithy** is a Ph.D. student in the University of Bridgeport pursuing his degree in Computer Science and Engineering. Abdelrahman has received his B.S. in Computer Science and MS in Computer Science in 2007 and 2008 respectively from the University of Bridgeport. Abdelrahman has participated twice in a sponsored program by the National Science Foundation, and the National Institutes of Health where he designed micro-electro-mechanical systems (MEMS) for biomedical applications. Current research interests of Abdelrahman include wireless communications, mobile communications and wireless sensor networks. 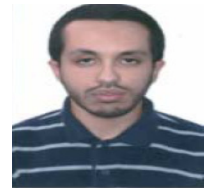

**Gonhsin Liu** is an associate professor in the Depratment of Computer Sceince and Engineering at the University of Bridgeport. Prof. Liu has received his B.S.from Tatung Institute of Technology and his M.S.and Ph.D.from State University of New York at Buffalo. Prof. Liu has research interests in Digital Signal, Image Processing, and Unix/Linux opertaing ssytems. 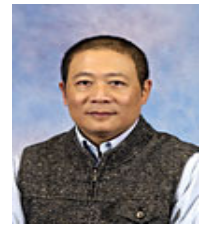